\documentclass[12pt]{article}
\usepackage{amsmath}
\numberwithin{equation}{section}
\title{Human migration and  the motion of substance in a channel of a network}
\author{Nikolay K. Vitanov$^{1,2}$, Kaloyan N. Vitanov$^1$}
\date{$^1$ Institute of Mechanics, Bulgarian Academy of Sciences, Acad. G. Bonchev Str., Bl. 4, 1113 Sofia, Bulgaria \\ 
$^{2}$ Max-Planck Institute for the Physics of Complex Systems, N{\"o}thnitzerstr. 38, 01187 Dresden, Germany}
\begin{document}
\maketitle
\begin{abstract}
We study the motion of a substance in a channel of a network that consists of  
chain of nodes of a network (the nodes can be considered as boxes) and edges 
that connect the nodes and form the way for motion of the substance. The nodes of the  
channel can have different "leakage", i.e., some amount of the substance can leave the 
channel at  a node and the rate of leaving may be  different for the different nodes of the 
channel.  In addition the nodes close to the end of the channel for some (construction or 
other) reason may be more "attractive" for the substance in comparison to the nodes  around 
the entry node of the channel. We discuss channels containing infinite  or finite number of nodes and obtain the distribution of the substance along the nodes. Two regimes of 
functioning of the channels are studied: stationary regime and non-stationary regime. 
The distribution of the substance along the nodes of the channel for the case of
stationary regime is a generalization of the Waring distribution (for channel with infinite 
number of nodes) or generalization of the truncated Waring distribution 
(for channel with finite number of nodes). In the non-stationary regime of 
functioning of the channel  one observes an exponential increase or exponential decrease of 
the amount of substance in the nodes. Despite this  the asymptotic distribution 
of the substance among the nodes of the channel in this regime is  stationary. 
The developed theory is applied for a study of the distrribution of migrants in countries that form migration
channels.
\end{abstract}
\section{Introduction}
In the last decades the researchers realized the importance of  dynamics of complex systems and this leaded to intensive studies of such systems, especially in the area of social dynamics and population dynamics \cite{cs1} - \cite{cs5}. In the course of these studies the networks have
appeared as important part of the structure of many complex systems \cite{cs6} - \cite{cs8}.

Research on network flows  has some of its roots in the studies on transportation 
problems, e.g.,  in the developing of minimal cost transportation models. This research 
topic was established in 1960's especially after the publishing the book of Ford and
Fulkerson \cite{ff}. At the beginning of the research the problems of interest
have been, e.g., how by minimal number of individuals to meet a fixed schedule 
of tasks; minimal cost flow problems; or possible maximal flows in a network.
In course of the years the area of problems connected to network flows has
increased very much. Today one uses the methodology from the theory of network
flows \cite{ahuja}, \cite{ch1} to solve problems connected to: (i) shortest 
path finding, (ii) just in time scheduling, (iii) facility layout and location, 
(iv) project management (determining minimum project duration), (v) optimal 
electronic route guidance in urban traffic networks \cite{hani}, (vi) 
self-organizing network flows, (vii) modeling and optimization of scalar
flows in networks \cite{ambro}, (viii) memory effects \cite{rosvall}, (ix) 
isoform identification of RNA \cite{bernard}, etc. (just some other examples 
are \cite{gomori} - \cite{boz}).
\par
Below we shall consider a specific network flow problem: motion of a substance
through a network channel in presence of possibility for "leakage" in the
nodes of the channel (loss of substance or usage of a part of substance in 
some process). In addition we shall assume the existence of possibility that 
the substance may have preference for some of the nodes of the channel (e.g.,
the channel may be structured in such a way that the substance tends to
concentrate in some of the nodes). This feature will allow us to use the model 
for study of motions of animals or humans. We note that the discussed model 
contains also the particular case when there is no preference of the substance 
with respect to the nodes of the channel. The obvious application of the model 
is for the flow of some non-living substance through a chanel with usage of part 
of substance for some industrial process in the nodes of the channel. We shall 
show that the model has more applications by another illustration: modeling of 
large human  migration flows. The large flows allow continuous modeling as in 
this case the discrete quantities can be approximated by a continuous ones. 
This choice of an illustration of the model has been made because of the
actuality of the problem of human migration  \cite{everet}. Indeed the study
of international migration becomes very actual after the large migration  
flows directed to  Europe in  2015. Much efforts are invested also in the 
study of internal migration in order to understand this migration and to
make projection of the  migration flows that may be very important for
taking decisions about economic development of regions of a country \cite{armi}-
\cite{borj}. Human migration models are of interest also for applied mathematics
as they can be classified as probability models (exponential model, Poisson model, multinomial model, Markov chain models of migration \cite{will99}- \cite{brl08}) or deterministic models (e.g., gravity model of migration \cite{grd05}). Human migration is closely connected to 
migration networks \cite{fawcet}, \cite{gurak}, to ideological struggles 
\cite{vit1},\cite{vit2}  and to waves and statistical distributions in 
population systems \cite{vit3} -\cite{vit6}. 
\par 
The paper is organized as follows. In Sect.2 we discuss a model for motion of
substance in a channel containing an infinite number of nodes. Two regimes of functioning 
of the channel: stationary regime  and non-stationary regime  are studied. 
Statistical distributions of the 
amount of substance in the nodes of the channel are obtained. A particular case
of the distribution for the stationary regime of functioning of the channel is 
the Waring distribution. Sect. 3 is devoted to the case of channel containing
finite number of nodes. This case is of interest for the problem of human
migration channels. The distribution of substance for stationary 
regime of functioning of such a channel is given by a distribution that is a  
generalization of the truncated
Waring distribution and a more complicated distribution is obtained for the
asymptotic state of the channel functioning in a non-stationary regime. In 
Sect. 4 we discuss application of the obtained mathematical results
to a human migration channel of finite length and discuss effects such as the 
possibility of concentration of migrants in the last node of the channel (the 
final destination country). Several concluding remarks are summarized in Sect. 5. 
\section{Channel containing infinite number of nodes}
We consider a channel consisting of a chain of nodes of a network.The nodes are
connected by edges and each node is connected only to the two neighboring 
nodes of the channel exclusive for
the first and the last node of the channel that are connected only to the
neighboring node. We study a model of the motion of substance through such a
channel which is an extension of the model discussed in  \cite{sg1} and
\cite{vk}. We consider each node as a cell (box), 
i.e.,  we consider an  array of infinite  number of cells indexed in 
succession by non-negative integers. The first cell has index $0$ and the last
cell has index $N$ (in the case discussed here $N=\infty$). 
We assume that an amount $x$ of some substance  is
distributed among the cells and this substance can move from one cell to another cell. Let $x_i$ be the amount of the substance in the $i$-th cell. Then
\begin{equation}\label{warigx1}
x = \sum \limits_{i=0}^\infty x_ i
\end{equation}
The fractions $y_i = x_i/x$ can be considered as probability values of
distribution of a discrete random variable $\zeta$
\begin{equation}\label{warigx2}
y_i = p(\zeta = i), \ i=0,1, \dots
\end{equation}
The content $x_i$ of any cell may change due to the following 3 processes:
\begin{enumerate}
\item Some amount $s$ of the substance $x$  enters the system of
cells from the external environment through the $0$-th cell;
\item Rate $f_i$ from $x_i$ is transferred from the $i$-th
cell into the $i+1$-th cell;
\item Rate $g_i$ from  $x_i$  leaks out the $i$-th cell into the
external environment.
\end{enumerate}
We assume that the process of the motion of the substance is continuous in the
time. Then the process can be modeled mathematically by the system of ordinary
differential equations:
\begin{eqnarray} \label{warigx4}
\frac{dx_0}{dt} &=& s-f_0-g_0; \nonumber \\
\frac{dx_i}{dt} &=& f_{i-1} -f_i - g_i, \ i=1,2,\dots.
\end{eqnarray}
\par
There are  two regimes of functioning of the channel: stationary regime 
and non-stationary regime.
\subsection{Stationary regime of functioning of the channel}
In the stationary regime of the functioning of the channel 
$\frac{dx_i}{dt}=0$, $i=0,1,\dots$. Let us mark the quantities for the stationary case with 
$^*$. Then from Eqs.(\ref{warigx4}) one obtains 
\begin{equation}\label{st1}
f_0^*=s^*-g_0^*; \ \  f_i^*=f_{i-1}^*-g_i.
\end{equation}  
This result can be written also as
\begin{equation}\label{st2}
f_i^* = s^*- \sum \limits_{j=0}^i g_j^*
\end{equation}
Hence for the stationary case the situation in the channel is determined by
the quantities $s^*$ and $g_j^*$, $j=0,1,\dots$. 
In this paper we shall assume the following forms of the amount 
of the moving substances  in 
Eqs.(\ref{warigx4}) ($\alpha, \beta, \gamma_i, \sigma$ are constants)
\begin{eqnarray}\label{warigx5}
s &=& \sigma x_0 = \sigma_0; \ \ \sigma_0 > 0  \nonumber \\
f_i &=& (\alpha_i + \beta_i i) x_i; \ \ \ \alpha_i >0, \ \beta_i \ge 0 \to
\textrm{cumulative advantage of higher nodes} \nonumber \\
g_i &=& \gamma_i x_i; \ \ \ \gamma_i \ge 0 \to \textrm{non-uniform leakage 
in the nodes}
\end{eqnarray}
The rules (\ref{warigx5}) differ from the rules in \cite{sg1} in 3 points: 
\begin{enumerate}
\item $s$ is proportional to the substance in the $0$th node  (the amount of this substance is $x_0$). In
\cite{sg1} $s$ is proportional to the amount $x$ of the substance in the entire
channel $x = \sum \limits_{i=0}^N x_i$;
\item Leakage rate $\gamma_i$ is different for the different nodes. In 
\cite{sg1} the leakage rate is constant and equal to $\gamma$ for all nodes 
of the channel (i.e., there is uniform leakage in the nodes).
\item
Parameters $\alpha_i$ and $\beta_i$ are different for the different cells.
In \cite{sg1} these parameters are the same for all cells of the channel.
\end{enumerate}
\par
Substitution of Eqs.(\ref{warigx5}) in Eqs.(\ref{warigx4}) leads to the
relationships
\begin{eqnarray}\label{wsx10}
\frac{dx_0}{dt} &=& \sigma_0 x_0 - \alpha_0 x_0 - \gamma_0 x_0; \nonumber \\
\frac{dx_i}{dt} &=& [\alpha_{i-1} + (i-1) \beta_{i-1}]x_{i-1} - (\alpha_i + i \beta_i  + \gamma_i)x_i; \ \ 
\ i=1,2,\dots
\end{eqnarray}
As we shall consider the stationary regime of functioning of the channel then
from the first of the Eqs.(\ref{wsx10}) it follows that
$\sigma_0 = \alpha_0 + \gamma_0$. This means that $x_0$
(the amount of the substance in the $0$-th cell of the channel) is free
parameter. In this case the solution of Eqs.(\ref{wsx10}) is
\begin{equation}\label{wsx11}
x_i = x_i^* + \sum \limits_{j=0}^i b_{ij} \exp[-(\alpha_j + j \beta_j + \gamma_j)t]
\end{equation}
where $x_i^*$ is the stationary part of the solution. For $x_i^*$ one obtains
the relationship
\begin{equation}\label{wsx12}
x_i^* = \frac{\alpha_{i-1} + (i-1) \beta_{i-1}}{\alpha_i + i \beta_i + \gamma_i} x_{i-1}^*, \ i=1,2,\dots
\end{equation}
The corresponding relationships for the coefficients $b_{ij}$ are
($i=1,\dots$):
\begin{equation}\label{wsx13}
b_{ij} = \frac{\alpha_{i-1} + (i-1) \beta_{i-1}}{(\alpha_i - \alpha_j) + (i \beta_i -
j \beta_j) + (\gamma_i - \gamma_j)} b_{i-1,j},
\ j=0,1,\dots,i-1
\end{equation}
From Eq.(\ref{wsx12}) one obtains
\begin{equation}\label{distr1}
x_i^* = \frac{\prod \limits_{j=0}^{i-1}[\alpha_{i-j-1}+(i-j-1)\beta_{i-j-1}]}{
\prod \limits_{j=0}^{i-1} \alpha_{i-j} + (i-j) \beta_{i-j} + \gamma_{i-j}} x_0^*
\end{equation}
The form of the corresponding stationary distribution $y_i^* = x_i^*/x^*$ 
(where $x^*$ is the amount of the substance in all of the cells of the channel)
is
\begin{equation}\label{distr}
y_i^* = \frac{\prod \limits_{j=0}^{i-1}[\alpha_{i-j-1}+(i-j-1)\beta_{i-j-1}]}{
\prod \limits_{j=0}^{i-1} \alpha_{i-j} + (i-j) \beta_{i-j} + \gamma_{i-j}} y_0^*
\end{equation}
(Note that the requirement $\sum \limits_{i=1}^{\infty} y_i^* =1$ has 
to be satisfied). To the best of our knowledge the distribution presented by
Eq.(\ref{distr}) was not discussed by other authors. Let us show that this
distributions contains as particular cases several famous distributions such as
Waring distribution, Zipf distribution, and Yule-Simon distribution. In order to
do this we consider the particular case when $\beta_i \ne 0$ and write $x_i$
from Eq.(\ref{distr1}) as follows
\begin{equation}\label{gen1}
x_i^* = \frac{\prod \limits_{j=0}^{i-1} b_{i-j} [k_{i-j-1} + (i-j-1)]}{\prod \limits_{j=0}^{i-1} [k_{i-j} + a_{i-j} + (i-j)]} x_0^*
\end{equation}
where $k_i = \alpha_i/\beta_i$; $a_i = \gamma_i/\beta_i$; 
$b_i = \beta_{i-1}/\beta_i$.
The form of the corresponding stationary distribution $y_i^* = x_i^*/x^*$ is
\begin{equation}\label{gen2}
y_i^* = \frac{\prod \limits_{j=0}^{i-1} b_{i-j} [k_{i-j-1} + (i-j-1)]}{\prod \limits_{j=0}^{i-1} [k_{i-j} + a_{i-j} + (i-j)]} y_0^*
\end{equation}
Let us now consider the particular case where $\alpha_i = \alpha$ and 
$\beta_i = \beta$ for $i=0,1,2,\dots$. Then from Eqs.(\ref{gen1}) and (\ref{gen2}) one obtains
\begin{equation}\label{wsx14}
x_i^* = \frac{[k+(i-1)]!}{(k-1)! \prod \limits_{j=1}^i (k+j+a_j)} x_0^*
\end{equation}
where $k = \alpha/\beta$ and $a_j=\gamma_j/\beta$.
The form of the corresponding stationary distribution $y_i^* = x_i^*/x^*$ is
\begin{equation}\label{wsx15}
y_i^* = \frac{[k+(i-1)]!}{(k-1)! \prod \limits_{j=1}^i (k+j+a_j)} y_0^*
\end{equation}
Let us consider the particular case where $a_0 = \dots = a_N$. In this case the
distribution from Eq.(\ref{wsx15}) is reduced to the distribution:
\begin{eqnarray}\label{wsx16}
P(\zeta = i) &=& P(\zeta=0) \frac{(k-1)^{[i]}}{(a+k)^{[i]}}; \ \ k^{[i]} = \frac{(k+i)!}{k!}; \ i=1, 2, \dots 
\end{eqnarray}
$P(\zeta=0)=y_0^* = x_0^*/x^*$ is the percentage of substance that is located in
the first cell of the channel. Let this percentage be 
\begin{equation}\label{wsx17}
y_0^* = \frac{a}{a+k}
\end{equation}
The case described by Eq.(\ref{wsx16}) corresponds to the situation where the
amount of substance in the first cell is proportional of the amount of substance
in the entire channel (self-reproduction property of the substance). In this
case Eq.(\ref{wsx15}) is reduced to the  distribution:
\begin{eqnarray}\label{wsx18}
P(\zeta = i) &=& \frac{a}{a+k} \frac{(k-1)^{[i]}}{(a+k)^{[i]}}; \ \ k^{[i]} = \frac{(k+i)!}{k!}; \ i=1, 2, \dots 
\end{eqnarray}
Let us denote $\rho = a$ and $k=l$. The distribution (\ref{wsx18}) is
exactly the Waring distribution (probability distribution of non-negative 
integers named after Edward Waring - the 6th Lucasian professor of Mathematics 
in Cambridge from the 18th century)  \cite{varyu1} - \cite{varyu3}
\begin{equation}\label{ap1}
p_l = \rho \frac{\alpha_{(l)}}{(\rho + \alpha)_{(l+1)}}; \
\alpha_{(l)} = \alpha (\alpha+1) \dots (\alpha+l-1)
\end{equation}
Waring distribution may be written also as follows
\begin{eqnarray}\label{ap2}
p_0 &=& \rho \frac{\alpha_{(0)}}{(\rho + \alpha)_{(1)}} = \frac{\rho}{\alpha + \rho}
\nonumber \\
p_l &=& \frac{\alpha+(l-1)}{\alpha+ \rho + l}p_{l-1}.
\end{eqnarray}
The mean $\mu$ (the expected value) of the Waring distribution is
\begin{equation}\label{ap3}
\mu = \frac{\alpha}{\rho -1} \ \textrm{if} \ \rho >1
\end{equation}
The variance of the Waring distribution is
\begin{equation}\label{ap4}
V = \frac{\alpha \rho (\alpha + \rho -1)}{(\rho-1)^2(\rho - 2)} \
\textrm{if} \ \rho >2
\end{equation}
$\rho$ is called the tail parameter as it controls the tail of the Waring
distribution. Waring distribution contains various distributions as particular cases. Let $i \to \infty$ Then the Waring distribution is reduced to  
\begin{equation}\label{ap5}
p_l \approx \frac{1}{l^{(1+\rho)}}.
\end{equation}
which is the frequency form of the Zipf distribution \cite{chen}.
If $\alpha \to 0$ the Waring distribution is reduced to the Yule-Simon distribution \cite{simon} 
\begin{equation}\label{ap6}
p(\zeta = l \mid \zeta > 0) = \rho B(\rho+1,l)
\end{equation}
where $B$ is the beta-function. 
\subsection{Non-stationary regime of functioning of the channel}
In the nonstationary case $dx_0/dt \ne 0$. In this case the solution of the
first equation of the system of equations (\ref{wsx10}) is
\begin{eqnarray}\label{nst11}
x_0 = b_{00} \exp[(\sigma_0 - \alpha_0 - \gamma_0)t] 
\end{eqnarray}
where $b_{00}$ is a constant of integration. $x_i$ must be obtained by solution of the corresponding  Eqs.(\ref{wsx10}). The form of $x_i$ is
\begin{eqnarray}\label{nst12}
x_i = \sum \limits_{j=0}^i b_{ij} \exp[-(\alpha_j + j \beta_j + \gamma_j - \sigma_j )t]
\end{eqnarray}
The solution of the system of equations (\ref{wsx10}) is (\ref{nst12})
where $\sigma_i =0$, $i=1,\dots,$ ($\sigma_0 = \sigma$): 
\begin{eqnarray}\label{nst13}
b_{ij} &=& \frac{\alpha_{i-1} + (i-1) \beta_{i-1}}{(\alpha_i - \alpha_j) +
(i \beta_i - j \beta_j) + (\gamma_i - \gamma_j)} b_{i-1,j}; \ i=1,\dots;
j=1,\dots, i-1 
\end{eqnarray}
and $b_{ii}$ are determined from the initial conditions in the cells of the
channel. 
The asymptotic solution ($t \to \infty$) is 
\begin{equation}\label{nst13}
x_i^a = b_{i0} \exp[(\sigma_0 - \alpha_0 - \gamma_0)t]
\end{equation}
This means that the asymptotic distribution $y_i^a = x_i^a/x^a$ is stationary
\begin{equation}\label{nst14}
y_i^a = \frac{b_{i0}}{\sum \limits_{j=0}^\infty b_{j0}}
\end{equation}
regardless of the fact that the amount of substance in the two cells may increase or decrease exponentially. 
The explicit form of this distribution is 
\begin{eqnarray}\label{nst15}
y_0^a &=& \frac{1}{1+ \sum \limits_{i=1}^{\infty} \prod \limits_{k=1}^i
\frac{\alpha_{k-1} + (k-1) \beta_{k-1}}{(\alpha_k - \alpha_0) + k \beta_k +(\gamma_k - \gamma_0)}} \nonumber \\
y_i^a &=& \frac{\prod \limits_{k=1}^i \frac{\alpha_{k-1} + (k-1) \beta_{k-1}}{(\alpha_k - \alpha_0) + k \beta_k + (\gamma_k - \gamma_0)}}{\sum \limits_{i=0}^{\infty} \prod \limits_{k=1}^i
\frac{\alpha_{k-1} + (k-1) \beta_{k-1}}{(\alpha_k - \alpha_0) + k \beta_k + (\gamma_k - \gamma_0)}}, 
i=1,\dots,
\end{eqnarray}
\section{Channel containing finite number of nodes}
Finite size channels are very interesting from the point of view of the
applications of the theory, e.g., to migrant flows.
Let us consider a channel consisting of   $N+1$ nodes (cells) and corresponding edges. The nodes are indexed in succession by
non-negative integers, i.e., the first cell has index $0$ and the last
cell has index $N$. In this case the total amount of substance
in the channel is
\begin{equation}\label{warig1}
x = \sum \limits_{i=0}^N x_ i
\end{equation}
The fractions $y_i = x_i/x$ can be considered as probability values of
distribution of a discrete random variable $\zeta$
\begin{equation}\label{warig2}
y_i = p(\zeta = i), \ i=0,1, \dots, N
\end{equation}
The mathematical model of the finite channel is as follows:
\begin{eqnarray} \label{warig4}
\frac{dx_0}{dt} &=& s-f_0-g_0; \nonumber \\
\frac{dx_i}{dt} &=& f_{i-1} -f_i - g_i, \ i=1,2,\dots, N-1 \nonumber \\
\frac{dx_N}{dt} &=& f_{N-1}  - g_N .
\end{eqnarray}
The relationships for the amount of the moving substances 
are the same as in the case of infinite channel:
\begin{eqnarray}\label{warigy5}
s &=& \sigma x_0 = \sigma_0 x_0; \ \ \sigma = \sigma_0 > 0  \nonumber \\
f_i &=& (\alpha_i + \beta_i i) x_i; \ \ \ \alpha_i >0, \ \beta_i \ge 0 \to
\textrm{cumulative advantage of higher nodes} \nonumber \\
g_i &=& \gamma_i x_i; \ \ \ \gamma_i \ge 0 \to \textrm{non-uniform leakage 
in the nodes}
\end{eqnarray}
Substitution of Eqs.(\ref{warigy5}) in Eqs.(\ref{warig4}) leads to the
relationships
\begin{eqnarray}\label{warig6}
\frac{dx_0}{dt} &=& \sigma_0 x_0 - \alpha_0 x_0 - \gamma_0 x_0; \nonumber \\
\frac{dx_i}{dt} &=& [\alpha_{i-1} + (i-1) \beta_{i-1}]x_{i-1} - (\alpha_i + i \beta_i + \gamma_i)x_i,
\ i=1,2,\dots,N-1
\nonumber \\
\frac{dx_N}{dt} &=& [\alpha_{N-1} + (N-1) \beta_{N-1}]x_{N-1} - \gamma_N x_N 
\end{eqnarray}
\subsection{Stationary regime of functioning of the channel}
In the stationary regime of functioning of the channel $\frac{dx_0}{dt}=0$, i.e. $\sigma_0 = \alpha_0 +  \gamma_0$.
In this case the system of equations (\ref{warig6}) has a stationary solution with
a free parameter $x_0$. This solution is
\begin{eqnarray}\label{ws1}
x_i^* &=& \frac{\alpha_{i-1} + (i-1) \beta_{i-1}}{\alpha_i + i \beta_i + \gamma_i} x_{i-1}^*, \ 
i=1,2,\dots, N-1 \nonumber \\
x_N^* &=& \frac{\alpha_{N-1} + (N-1) \beta_{N-1}}{\gamma_N} x_{N-1}^*.
\end{eqnarray}
The solution of Eqs.(\ref{warig6}) is 
\begin{equation}\label{ws2}
x_i = x_i^* + \sum \limits_{j=0}^i b_{ij} \exp[-(\alpha_j + j \beta_j + \gamma_j)t]
\end{equation}
The substitution of Eq.(\ref{ws2}) in Eqs.(\ref{warig6}) leads to the following
relationships for the coefficients $b_{ij}$ ($\alpha_N = \beta_N =0$ as there 
is no $N+1$-st node(cell) where the substance can move from the $N$-th mode
(cell))
\begin{eqnarray}\label{soly}
b_{ij} &=& \frac{\alpha_{i-1} + (i-1) \beta_{i-1}}{(\alpha_i - \alpha_j) + (i \beta_i - j \beta_j) + (\gamma_i - \gamma_j)} b_{i-1,j}; \ i=1,\dots,N-1
\nonumber \\
b_{Nj} &=&  \frac{\alpha_{N-1} + (N-1) \beta_{N-1}}{\gamma_N - \gamma_j - \alpha_j - j \beta_j} b_{N-1,j}, \ j=0, \dots, N-1
\end{eqnarray}
$b_{ij}$ that are not determined by Eqs.(\ref{soly}) may be determined by the 
initial conditions and in this process $b_{00}$ may be fixed too. 
In the exponential
function in Eq.(\ref{ws2}) there are no negative coefficients and because of this 
when $t \to \infty$   the system comes to the
stationary solution from Eqs.(\ref{ws1}). The form of this stationary solution
is
\begin{eqnarray}\label{ws3}
x_i^* &=& \frac{\prod \limits_{j=1}^i [\alpha_{i-j} + (i-j) \beta_{i-j}] }{
\prod \limits_{j=1}^i (\alpha_j + j \beta_j + \gamma_j)} x_0^*, \
i=1,\dots, N-1 \nonumber\\
x_N^* &=& \frac{\prod \limits_{j=1}^N [\alpha_{N-j} + (N-j) \beta_{N-j}]}{ \gamma_N \prod \limits_{j=1}^{N-1} (\alpha_j + j \beta_j + \gamma_j)} x_0^*
\end{eqnarray}
Let $x^*=\sum \limits_{i=0}^N x_i$ be the total amount of the substance for 
the case of stationary state of the channel. Then we can consider the \
distribution $y_i^* = x_i^*/x^*$. Its form is
 \begin{eqnarray}\label{ws4}
y_i^* &=& \frac{\prod \limits_{j=1}^i [\alpha_{i-j} + (i-j) \beta_{i-j}] }{
\prod \limits_{j=1}^i (\alpha_j + j \beta_j + \gamma_j)} y_0^*, \
i=1,\dots, N-1 \nonumber\\
y_N^* &=& \frac{\prod \limits_{j=1}^N [\alpha_{N-j} + (N-j) \beta_{N-j}]}{ \gamma_N \prod \limits_{j=1}^{N-1} (\alpha_j + j \beta_j + \gamma_j)} y_0^*
\end{eqnarray}
To the best of our knowledge the distribution presented by
Eq.(\ref{ws4}) was not discussed by other authors.
Let us consider the particular case where $\alpha_i = \alpha$, $\beta_i = \beta$ and $\gamma_i = \gamma$ and $k = \alpha/\beta$, $a = \gamma/\beta$. 
In this case the
distribution from Eq.(\ref{ws4}) is reduced to the distribution:
\begin{eqnarray}\label{ws5}
P(\zeta = i) &=& P(\zeta=0) \frac{(k-1)^{[i]}}{(a+k)^{[i]}}; \ \ k^{[i]} = \frac{(k+i)!}{k!}; \ i=0,\dots,N-1 \nonumber \\
P(\zeta = N) &=& \frac{P(\zeta=0)}{a} \frac{(k-1)^{[N]}}{(a+k)^{[N-1]}},
\end{eqnarray}
$P(\zeta=0)=y_0^* = x_0^*/x^*$ is the percentage of substance that is located in
the first node of the channel. Let this percentage be 
\begin{equation}\label{ws6}
y_0^* = \frac{a}{a+k}
\end{equation}
The case described by Eq.(\ref{ws6}) corresponds to the situation where the
amount of substance in the first node is proportional of the amount of substance
in the entire channel (self-reproduction property of the substance). In this
case Eq.(\ref{ws5}) is reduced to the truncated Waring distribution:
\begin{eqnarray}\label{warig12}
P(\zeta = i) &=& \frac{a}{a+k} \frac{(k-1)^{[i]}}{(a+k)^{[i]}}; \ \ k^{[i]} = \frac{(k+i)!}{k!}; \ i=0,\dots,N-1 \nonumber \\
P(\zeta = N) &=& \frac{1}{a+k} \frac{(k-1)^{[N]}}{(a+k)^{[N-1]}},
\end{eqnarray}
\par
The truncated Waring distribution (\ref{warig12}) is close to the 
Waring distribution that was discussed above in the text.
A characteristic feature of the truncated Waring
distribution is the possibility for accumulation of substance in the last
node  of the channel (and this concentration can be quite significant)
\cite{vk}.
\par
Let us note that that for the case of distribution (\ref{ws4}) $y_N^* > y_{N-1}^*$ when $k+(N-1) > a_N$, i.e., the concentration of the substrate in
the last node of the channel depends on the situation in the last two nodes
(from the parameters $a_{N-1}, \beta_{n-1}, \gamma_N$). For the case of truncated Waring distribution (\ref{warig12})
$y_N^* > y_{N-1}^*$ when $k+(N-1) > a$ where $a= \sigma/\beta$,i.e., the concentration of substance in the last node of the channel depends on the
situation in the first node of the channel. This may be important for the 
case of channels for human migrants.
\subsection{Non-stationary regime of functioning of the channel}
In the nonstationary case $dx_0/dt \ne 0$. In this case the solution of the
first equation of the system of equations (\ref{warig6}) is
\begin{eqnarray}\label{ws7}
x_0 = b_{00} \exp[(\sigma_0 - \alpha_0 - \gamma_0)t] 
\end{eqnarray}
where $b_{00}$ is a constant of integration. $x_i$ must be obtained by solution of the corresponding  Eqs.(\ref{warig6}). The form of $x_i$ is
\begin{eqnarray}\label{ws8}
x_i = \sum \limits_{j=0}^i b_{ij} \exp[-(\alpha_j + j \beta_j + \gamma_j - \sigma_j )t]
\end{eqnarray}
In order to understand the processes in the channel let us consider first the
case of channel consisting of two nodes ($N=1$). In this case we have to solve the additional equation
\begin{equation}\label{nsc1}
\frac{dx_1}{dt} = \alpha_0 x_0 - \gamma_1 x_1
\end{equation}
The solution is
\begin{equation}\label{nsc2}
x_1 = \frac{\alpha_0}{\gamma_1 - \gamma_0 + \sigma_0 - \alpha_0} b_{00} \exp[(\sigma_0 - \alpha_0 - \gamma_0)t] + b_{11} \exp(-\gamma_1 t)
\end{equation}
$b_{11}$ can be determined from the initial conditions at $t=0$. The  asymptotic form of the obtained solution ($t \to \infty$) is
\begin{eqnarray}\label{nsc3}
x_0^a &=& b_{00} \exp[(\sigma_0 - \alpha_0 - \gamma_0)t] \nonumber \\
x_1^a &=& \frac{\alpha_0}{\gamma_1 - \gamma_0 + \sigma_0 - \alpha_0} b_{00} \exp[(\sigma_0 - \alpha_0 - \gamma_0)t]
\end{eqnarray}
as $\gamma_1 >0$. Let us consider the asymptotic distribution $y_i^a = x_i^a/x^a$ where $x^a = \sum \limits_{i=0}^N x_i$. $x_0^a$ and $x_1^a$ depend on
$t$ but nevertheless the asymptotic distribution is stationary
\begin{equation}\label{nsc4}
y_0^a = \frac{1}{1+ \frac{\alpha_0}{\gamma_1 - \gamma_0 + \sigma_0 - \alpha_0}}; \ \
y_1^a = \frac{1}{1+ \frac{\gamma_1 - \gamma_0 + \sigma_0 - \alpha_0}{\alpha_0}}
\end{equation}
Thus the distribution of the substance in the channel tend to a stationary
asymptotic distribution regardless of the fact that the amount of substance in
the two nodes may increase or decrease exponentially. 
\par
Let us now consider the case of channel containing more than 2 nodes ($N>1$). In this case the solution of the system of equations (\ref{warig6}) is (\ref{ws8})
where $\sigma_i =0$, $i=1,\dots,N-1$; 
\begin{eqnarray}\label{ws9}
b_{ij} &=& \frac{\alpha_{i-1} + (i-1)\beta_{i-1}}{(\alpha_i - \alpha_j) +
(i\beta_i - j\beta_j) +(\gamma_i - \gamma_j)} b_{i-1,j}; \ i=1,\dots,N-1
\nonumber \\
b_{Nj} &=&  \frac{\alpha_{N-1} + (N-1) \beta_{N-1}}{ \gamma_N - \gamma_j - \alpha_j - j \beta_j} b_{N-1,j}, \ j=0, \dots, N-1
\nonumber \\
\sigma_N &=& \alpha_N + N \beta_N, 
\end{eqnarray}
and $b_{ii}$ are determined from the initial conditions in the nodes of the
channel. The asymptotic solution ($t \to \infty$) is 
\begin{equation}\label{nsc5}
x_i^a = b_{i0} \exp[(\sigma_0 - \alpha_0 - \gamma_0)t]
\end{equation}
This means that the asymptotic distribution $y_i^a = x_i^a/x^a$ is stationary
\begin{equation}\label{ncs6}
y_i^a = \frac{b_{i0}}{\sum \limits_{j=0}^N b_{j0}}
\end{equation}
regardless of the fact that the amount of substance in the two cells many increase or decrease exponentially. The explicit form of this distribution is 
\begin{eqnarray}\label{nsc7}
y_0^a &=& \frac{1}{\Omega} \nonumber \\
y_i^a &=& \frac{\prod \limits_{k=1}^i
\frac{\alpha_{i-k} + (i-k) \beta_{i-k}}{(-\alpha_0 + \alpha_{i-k+1}) +
(i-k+1)\beta_{i-k+1} + (-\gamma_0 + \gamma_{i-k+1})}}
{\Omega},
\nonumber \\
i&=&1,\dots, N-1 \nonumber \\
y_N^a &= &  \frac{\frac{\alpha_{N-1} + (N-1) \beta_{N-1}}{(\gamma_N - \gamma_0) - \alpha_0} \prod \limits_{k=1}^{N-1}
\frac{\alpha_{i-k} + (i-k) \beta_{i-k}}{(-\alpha_0 + \alpha_{i-k+1}) +
(i-k+1)\beta_{i-k+1} + (-\gamma_0 + \gamma_{i-k+1})}}{\Omega}\nonumber \\
\end{eqnarray}
where
\begin{eqnarray}\label{omg}
\Omega &=& 1+ \sum \limits_{i=1}^{N-1} \prod \limits_{k=1}^i
\frac{\alpha_{i-k} + (i-k) \beta_{i-k}}{(-\alpha_0 + \alpha_{i-k+1}) +
(i-k+1)\beta_{i-k+1} + (-\gamma_0 + \gamma_{i-k+1})} + \nonumber \\
&& \frac{\alpha_{N-1} + (N-1) \beta_{N-1}}{(\gamma_N - \gamma_0) - \alpha_0} \prod \limits_{k=1}^{N-1}
\frac{\alpha_{i-k} + (i-k) \beta_{i-k}}{(-\alpha_0 + \alpha_{i-k+1}) +
(i-k+1)\beta_{i-k+1} + (-\gamma_0 + \gamma_{i-k+1})} \nonumber \\
\end{eqnarray}
\section{Application of obtained results to channels of migration networks}
The model discussed above can be used for a study of motion of substance
through cells of appropriate technological systems. The model can be applied 
also for investigation of other systems. Below we shall discuss it in connection with channels of human migration for the case when the migration flows are large and
continuous approximation of these flows can be used.
Let us consider a chain of $N+1$ countries or cities. 
This chain may be considered as a channel in  a migration network.
The nodes of this network (corresponding to the countries of the channel for an example) may be considered as boxes (cells). A flow of migrants moves through this migration channel from the country of entrance to the final destination country. The entry country will be the node with label $0$
and the final destination country will be the node with label $N$. 
Let us have  a number $x$ of migrants that are
distributed among the countries. Let $x_i$ be the number of migrants in the
$i$-th country. This number can change on the basis of the following three processes: (i) 
A number $s$ of migrants enter the channel from the external environment through 
the $0$-th node (country of entrance);
A rate $f_i$ from $x_i$ is  transferred from the $i$-th country  to the 
$i+1$-th country; (iii) A rate $g_i$ from $x_i$ change their status (e.g. 
they are not anymore
migrants and may become citizens of the corresponding country, may return home, etc.).
The values of $x_i$ can be determined by Eqs. (\ref{warig4}).
The relationships (\ref{warigx5}) mean that: (i) 
The number of migrants $s$ that 
enter the channel is proportional of the current number of migrants in the entry
country of the channel; (ii) There may be preference for some countries, e.g. migrants may prefer
the countries that are around the end of the migration channel (and the
final destination country may be the most preferred one); (iii)
It is assumed that the conditions along the channel are different with respect
to 'leakage' of migrants, e.g. the different rates $\gamma_i$ of migrants 
leave the flow of migrants in different countries of the channel. 
In addition the the transition from country to country may have different 
grade of difficulty (different
$\alpha_i$) and the attractiveness of the countries along the channel may be
different for migrants (different $\beta_i$). 
\par
$\sigma_0$ is the "gate" parameter as it regulates the number of migrants that enter the channel.
The parameters $\gamma_i$ regulate the "absorption" of the channel as they reflect 
the change of the status of some migrants. 
The large values of $\gamma_i$ may compensate the value of $\sigma$ and even may
lead to decrease of the number of migrants in the channel. The large values of
$\gamma_i$ may however lead to integration problems connected to migrants.  
\par
Small values of  parameters $\alpha_i$ mean that the way of the migrants through the channel is more difficult and because of this the migrants tend to concentrate in the entry country (and eventually in the second
country of the channel). The countries that are in the second half of the migration channel and especially the final destination country may try to decrease $\alpha_i$ by agreements that commit the entry country to keep the migrants on its territory. Any increase of $\alpha_i$ may lead to increase of 
the proportion of migrants that reach the second half of the migration channel and especially the final destination country.
\par
The parameters $\beta_i$ regulate the attractiveness of the countries along the migration channel. Large values of $\beta_i$ mean that the remaining countries
in the channel and especially the final destination country are very attractive 
for some reason. This increases the attractiveness of
the countries from the second half of the channel (migrants want more to reach these countries as in such a way the distance to the final destination country
is smaller). If for some reason $\beta_i$ are kept at high values a flood of migrants may reach the final destination country which may lead to large logistic and other problems. 
\par
Let us now consider the case of channel consisting of finite number of
nodes and the stationary case. Then from Eq.(\ref{ws1}) we obtain the relationship
\begin{equation}\label{m1}
\frac{y_N^*}{y_{N-1}^*} = \frac{\alpha_{N-1} + (N-1) \beta_{N-1}}{\gamma_N}
\end{equation}
If $\frac{y_N^*}{y_{N-1}^*} >1$ there is an effect of concentration of migrants
in the final destination country. This happens when
\begin{equation}\label{m2}
\alpha_{N-1} + (N-1) \beta_{N-1} > \gamma_N 
\end{equation}
i.e., if the attractivity of the final destination country is large (large
value of $\beta_{N-1}$) and it is relatively easy to cross the border to the final destination country (large $\alpha_{N-1}$) and the probability of change of the status of migrants in the final destination country of the
channel are not large enough to compensate for this popularity (value of
$\gamma_N$  is relatively small). In order to avoid the arising
of the effect of the concentration of migrants in the final destination country
one has to achieve
\begin{equation}\label{m3}
\alpha_{N-1} + (N-1)\beta_{N-1} < \gamma_N 
\end{equation}
This means that one should try to decrease the
number of migrants entering the final destination country (to lower the
value of $\alpha_{N-1}$); to decrease the attractivity of the final destination country (to lower the value of $\beta_{N-1}$) and to increase the probabilities for
change of the status of the migrants in the entry country and/or in the final
destination country (to increase the value of  $\gamma_N$).
\par
A new effect with respect to the theory developed in our previous study 
\cite{vk} is
the possibility of accumulation of migrants not only in the final destination 
country but also in any country of the channel. This can happen (see 
Eqs.(\ref{ws1})) when (for some value of $i$)
\begin{equation}\label{m3x}
\frac{\alpha_{i-1} + (i-1) \beta_{i-1}}{\alpha_i + i \beta_i + \gamma_i} >1
\end{equation}
Eq.(\ref{m3x}) shows that such a case of concentration of migrants may happen when the entry of
migrants in the $i$-th country is easy and this country is popular among
migrants (large values of $\alpha_{i-1}$ and $\beta_{i-1}$ with respect to
$\alpha_i$ and $\beta_i$) and in addition the value of $\gamma_N$ is small.
\par
The change of the total number of migrants in the channel can be obtained by
taking the sum of the equations (\ref{warig6}). The result is
\begin{equation}\label{m4}
\frac{dx}{dt}= \sigma_0 x_0 - \sum \limits_{l=0}^N \gamma_l x_l
\end{equation}
Thus the total number of migrants in the channel may increase fast when
many migrants enter the channel (when the value of $\sigma_0$ is large) and decrease when the probability for change of the status of the migrants increases
along the channel (when the values of  some of $\gamma_l$ or the values of 
all $\gamma_l$ increase). 
\par
In the stationary regime of functioning of a finite channel 
the total number of migrants in the countries of the channel is
\begin{equation}\label{m5}
x^* = x^*_0 \left \{1 + \sum \limits_{i=1}^{N-1} 
\frac{\prod \limits_{j=1}^i [\alpha_{i-j} + (i-j) \beta_{j-1}] }{
\prod \limits_{j=1}^i (\alpha_j + j \beta_j + \gamma_j)} +
\frac{\prod \limits_{j=1}^N [\alpha_{i-j} + (N-j) \beta_{j-1}]}{ \gamma_N \prod \limits_{j=1}^{N-1} (\alpha_j + j \beta_j + \gamma_j)}
  \right \} 
\end{equation}
Thus the number of migrants in the countries of the channel may be decreased if one manages 
to decrease the number of migrants in the entry country of the channel (i.e. if one manages 
to decrease the value of $x^*_0$). 
\par
Finally let us consider the case of channel consisting of finite number of 
countries and in stationary regime of functioning. Let in addition the
motion of migrants be determined  by the attractivity of the final destination 
country at the expense of the easiness of moving through the borders between the countries of the channel (i.e., $\beta_i >> \alpha_i$ and
parameter $\alpha_i$ may be neglected in the relationships for 
all cells except  in the
numerator for the $0$-th cell in Eq.(\ref{ws1})). Let in addition there be
no "leakage" (the migrants are very much fascinated by the final destination
countries and the change of the statuses along the channel is small, i.e.,
one can neglect all $\gamma_i$ except for $\gamma_N$ which is assumed to be significant).
Thus from Eq.(\ref{ws1}) we obtain
\begin{equation}\label{m8}
x_1^* = \frac{\alpha_0}{\beta_1} x_0^*; \ x_i^* = \frac{(i-1)}{ i} \frac{\beta_{i-1}}{\beta_i} x^*_{i-1}, i=2,..., N-1; \ x_N^* = \frac{ (N-1)\beta_{N - 1}}{\gamma_N} x_{N-1}^*.
\end{equation}
From the second of equations (\ref{m8}) one easily obtains
$x_i^* = \frac{i-k}{i} \frac{\beta_{i-k}}{\beta_i} x^*_{i-k}$. Then the approximate total number of the migrants in the channel is
\begin{eqnarray}\label{m9a}
x^* &=&  x_0^* \left \{ 1+ \alpha_0 \sum \limits_{l=1}^{N-1}
\frac{1}{l \beta_l} + \frac{\alpha_0}{\gamma_N} \right \} 
\end{eqnarray}
and the distribution of the migrants among the countries of the channel is
\begin{eqnarray}\label{gen_dis}
y_0^* &=& \frac{1}{1+ \alpha_0 \sum \limits_{l=1}^{N-1}
\frac{1}{l \beta_l} + \frac{\alpha_0}{\gamma_N}} \nonumber \\
y_1 &=& \frac{\alpha_0}{\beta_1 \left[1+ \alpha_0 \sum \limits_{l=1}^{N-1}
\frac{1}{l \beta_l} + \frac{\alpha_0}{\gamma_N} \right]} \nonumber \\
y_i^* &=& \frac{\alpha_0}{i \beta_i \left[ 1+ \alpha_0 \sum \limits_{l=1}^{N-1}
\frac{1}{l \beta_l} + \frac{\alpha_0}{\gamma_N}\right]},
\ \ i=2, \dots, N-1 \nonumber \\
y_N^* &=& \frac{\alpha_0}{\gamma_N \left[ 1+ \alpha_0 \sum \limits_{l=1}^{N-1}
\frac{1}{l \beta_l} + \frac{\alpha_0}{\gamma_N}\right]}
\end{eqnarray}
Let us now denote $\alpha_0$ as $\alpha$ and assume that $\beta_1 = \dots =
\beta_{N-1} = \beta$. Then
\begin{eqnarray}\label{m9}
x^* &=&  x_0^* \left \{ 1+ \frac{\alpha}{\beta} \sum \limits_{l=1}^{N-1}
\frac{1}{l} + \frac{\alpha}{\gamma_N} \right \} = x^*_0  \left \{ 
1 + \frac{\alpha}{\beta} H_{N-1} + \frac{\alpha}{\gamma} \right \}
\end{eqnarray}
where $H_{N-1}$ is the $N-1$-th harmonic number. Let us use the approximate
relationship for harmonic numbers $H_N = \ln(N) + C_E + \frac{1}{2N} - \frac{1}{12 N^2} + \frac{1}{120 N^4} - \dots$ ($C_E$ is the constant of Euler). Then for a migration chanel of finite and
not very large length we obtain the relationship
\begin{equation}\label{m10}
x^* \approx x^*_0  \left \{ 
1 + \frac{\alpha}{\beta} \left[ \ln(N-1) + C_E + \frac{1}{2(N-1)} \right] + \frac{\alpha}{\gamma} \right \}
\end{equation}
and the approximate distribution of the migrants in the channel $y_i^* = \frac{x_i^*}{x^*}$ will be
\begin{eqnarray}\label{m11}
y_0^* &=& \frac{1}{\left \{ 
1 + \frac{\alpha}{\beta} \left[ \ln(N-1) + C_E + \frac{1}{2(N-1)} \right] + \frac{\alpha}{\gamma} \right \}} \nonumber \\
y_1^* &=&  \frac{\alpha}{\beta  \left \{ 
1 + \frac{\alpha}{\beta} \left[ \ln(N-1) + C_E + \frac{1}{2(N-1)} \right] + \frac{\alpha}{\gamma} \right \}}\nonumber \\
y_i^* &=& \frac{\alpha}{\beta i \left \{ 
1 + \frac{\alpha}{\beta} \left[ \ln(N-1) + C_E + \frac{1}{2(N-1)} \right] + \frac{\alpha}{\gamma} \right \}}, i=2,\dots,N-1 \nonumber \\
y_N^* &=&  \frac{\alpha}{\gamma_N \left \{ 
1 + \frac{\alpha}{\beta} \left[ \ln(N-1) + C_E + \frac{1}{2(N-1)} \right] + \frac{\alpha}{\gamma} \right \}}
\end{eqnarray}
which is a version of truncated Zipf distribution. 
\par 
As an application of the above theory we shall discuss a stationary regime 
of functioning of a migration channel, containing 3 countries. Let the migrants 
come overseas (e.g., by boats) to the entry country of the channel. The attractive country in the channel is the third country of the channel (the final destination country).
The second country of the channel is not attractive for the migrants but in order to
reach the final destination country the migrants have to move through its 
territory. Let the flow of migrants
be large. Then we can apply the part of the model described by Eqs. (\ref{ws3})
and (\ref{ws4}). The number of the migrants in the three countries of the channel are 
\begin{eqnarray}\label{ex1}
x^*_0; \ \  x^*_1 = x^*_0 \frac{\alpha_0}{\alpha_1+\beta_1+\gamma_1}; \ \
x_2^* = x_0^* \frac{\alpha_1+\beta_1}{\gamma_2} \frac{\alpha_0}{\alpha_1+\beta_1 + \gamma_1}
\end{eqnarray}
The total number of migrants in the countries of the channel will be
\begin{equation}\label{ex2}
x^* = x_0^* \left( 1+ \frac{\alpha_0}{\alpha_1+\beta_1+\gamma_1} +
\frac{\alpha_1+\beta_1}{\gamma_2} \frac{\alpha_0}{\alpha_1+\beta_1 + \gamma_1} \right)
\end{equation}
and the distribution of the migrants along the countries of the channel will be
\begin{eqnarray}\label{ex3}
y_0^* = \frac{1}{\left( 1+ \frac{\alpha_0}{\alpha_1+\beta_1+\gamma_1} +
\frac{\alpha_1+\beta_1}{\gamma_2} \frac{\alpha_0}{\alpha_1+\beta_1 + \gamma_1} \right)}
\nonumber \\
y_1^* =\frac{\frac{\alpha_0}{\alpha_1+\beta_1+\gamma_1}}{\left( 1+ \frac{\alpha_0}{\alpha_1+\beta_1+\gamma_1} +
\frac{\alpha_1+\beta_1}{\gamma_2} \frac{\alpha_0}{\alpha_1+\beta_1 + \gamma_1} \right)} 
\nonumber \\
y_2^* =\frac{\frac{\alpha_1+\beta_1}{\gamma_2} \frac{\alpha_0}{\alpha_1+\beta_1 + \gamma_1}}{\left( 1+ \frac{\alpha_0}{\alpha_1+\beta_1+\gamma_1} +
\frac{\alpha_1+\beta_1}{\gamma_2} \frac{\alpha_0}{\alpha_1+\beta_1 + \gamma_1} \right)}
\end{eqnarray}
\par 
Let us discuss two scenarios. In the first scenario there are no measures to
decrease the number of migrants in the entry country of the channel and their
number remains, e.g., $x_0^* = 300,000$. let the stationary state of the channel
be characterized by values of parameters: $\alpha_0 = \alpha_1 = \beta_1 = \gamma_1 = \gamma_2 = 0.0001$.
Then in countries of the channel there are $600,000$ migrants: $300,000$ in the 
entry country ($50\%$), $100,000$ in the second country of the channel (about 
$16,7\%$) and $200,000$ in the final destination country (about $33.3\%$). Now let the number of the 
migrants in the first country of the channel is a large burden for this country,
and it  decides to increase $\alpha_0$ (e.g., to ease the border control and to increase the probability that the 
migrants may move successfully (and mostly illegally) to the second country of the 
channel. Let the result of such a behaviour be that $\alpha_0$ increase from $0.0001$ to
$0.0002$. After some time the channel will have new stationary state of 
operation where  $300,000$ ($x_0^*$ remains unchanged) but the number of
migrants in the second and third country of the channel will increase to
$200,000$ in the second country of the channel and $400,000$ in the final
destination country. Then the second country of  the channel may decide that 
the migration burden is too high for it and this country take measures that 
may lead to increase of $\alpha_1$. Let this increase be
from $\alpha_1 =0.0001$ to $\alpha_1=0.0002$. We remember that 
$\alpha_0=0.0002$ and the  other parameters of the channel have values of 
$0.0001$. This will have the following effect of the stationary regime of the
functioning of the channel: there will be $300,000$ migrants in the entry country
of the channel (as no measures are taken to decrease $x_0^*$). Then there will
be $150,000$ migrants in the second country of the channel (increasing value of
$\alpha_1$ decreased the number of migrants in this country) and there will be
$450,000$ migrants in the final destination country (part of the migrants move
from the second country of the channel to the final destination country). Let us
now the final destination country decides that the migration burden it too large
for it and takes measures to decrease its popularity. Let these measures lead to
a new value of $\beta_1$: $\beta_1 =0$. Then the stationary regime of the
functioning of the migration channel will be characterized by the following
number of migrants in the three countries: $300,000$ migrants in the entry
country, $200,000$ migrants in the second country of the channel, and $400,000$
migrants in the final destination country of the channel. Thus the number of
migrants in the final destination country will decrease at the expense of the
migrants in the second country of the channel.
\par 
The sole actions of the countries from the scenario 1 above can continue but as
we have seen this will not lead to significant decrease of migration burden.
Such significant decrease can be realized in the scenario 2: the countries
concentrate their efforts in order to decrease $x_0$ keeping the values
of the other parameters unchanged. This may have large effect as follows.
Let the initial stationary state of the channel be characterized by 
$x_0^*=300,000$, and $\alpha_0 = \alpha_1 = \beta_1 = \gamma_1 = 
\gamma_2 = 0.0001$. Then the number of migrants in the second country of the
channel will be $100,000$ and the number of migrants in the final destination
country will be $200,000$.  Let now the measures be taken and the number of
migrants in the entry country of the channel decreases to $x_0^*=200,000$. Then
the stationary state of the channel will be characterized by about $67,000$
migrants in the second country of the channel and about $133,000$ migrants
in the final destination country. If the measures for decreasing $x_0^*$
continue and the number of migrants reduces to $45,000$ then the corresponding
numbers of migrants will be $15,000$ in the second country of the channel and
$30,000$ in the final destination country. Thus scenario 2 is much more
effective from the point of view of decreasing migration burden on the countries
of the channel. 
\section{Concluding remarks}
Above we have discussed a model for motion of a substance in channels of
networks. Two regimes of functioning of the channel are studied: stationary
regime of motion of the substance and nonstationary regime of motion of the
substance. The main result of the study are the obtained distributions of substance
along the cells for the case of stationary regime of motion of the substance.
The corresponding distribution for the case of channel containing infinite
number of modes is a generalization of the Waring distribution. The distribution
obtained for the case of  a channel
containing finite number of nodes is a generalization of the truncated Waring
distribution. In addition to classical application of the model for calculation
of motion and distribution of substance in channels of technological systems we
have discussed the model also in the context of motion of large amounts of migrants through 
migration channels, e.g., connecting several countries. Specific
characteristic of the discussed model is the possibility for different 
"leakage" of the nodes, i.e., the different probabilities for a change of the
status of a migrant in the different nodes (countries) of the channel.
For the case of non-stationary regime of functioning of the channel  the 
number of migrants in the channel may increase or decrease exponentially 
but the asymptotic distribution of the migrants in the
countries of the channel is stationary and depends strongly on the situation 
in the first node (entry country) of the channel. The corresponding 
distributions of the migrants in the countries of the channel are more complicated in 
comparison to the distributions for the case of   stationary regime of
functioning of the channel. In the stationary regime of functioning of
a channel consisting of finite number of nodes an effect of concentration 
of migrants in the final destination country (last node of the channel) 
may be observed if the final destination country is popular enough.
The possibility of different "leakage"
and different preferences may lead to another concentration effect:  the 
concentration of migrants may happen not only in the last country of the 
channel but also in the countries that are between the entry country of the 
channel and final destination country. Let us note finally that if the 
popularity of the countries close to the final 
destination country is large and the motion from node to node (from 
country to country) is not so easy and in addition the migrants are 
not interested in change of their status in the countries of the 
channel except for the final destination country, then the 
distribution of the migrants in the channel is a version of the Zipf 
distribution.

\end{document}